\documentclass[prl,twocolumn,nofootinbib,
showpacs,superscriptaddress,preprintnumbers,amsmath,amssymb]{revtex4-1}

\usepackage{graphicx, color}
\usepackage{dcolumn}
\usepackage{bm}
\usepackage{amsmath}
\usepackage{amsthm}

\newcommand{\diff}{\mathrm{d}}

\newcommand\diag{\operatorname{diag}}

\newcommand\p{\partial}

\usepackage[normalem]{ulem}  % \sout{old text} for strikeout

\renewcommand\sout{\bgroup \color{red} \ULdepth=-.5ex \ULset}

\newcommand{\be}{\begin{equation}}
\newcommand{\ee}{\end{equation}}
\newcommand{\bea}{\begin{eqnarray}}
\newcommand{\eea}{\end{eqnarray}}
\newcommand{\ve}{\varepsilon}
\newcommand{\im}{\mathrm{i}}
\newcommand{\rmc}{\mathrm{c}}
\newcommand{\rme}{\mathrm{e}}
\newcommand{\rmf}{\mathrm{f}}

\usepackage[colorlinks=true, pdfstartview=FitV, linkcolor=red,
citecolor=blue, urlcolor=blue]{hyperref} 

\begin{document}
%%%%%%%%%%%%%%%%%%%%%% Title page %%%%%%%%%%%%%%%%%%%%%%%%%%%%%%

\title{
Quark-hadron continuity beyond Ginzburg-Landau paradigm
}

\author{Yuji~Hirono}
\email{yuji.hirono@apctp.org}

\affiliation{
Asia Pacific Center for Theoretical Physics, Pohang 37673, Korea
}
\affiliation{
Department of Physics, POSTECH, Pohang 37673, Korea
}

\author{Yuya Tanizaki}
\email{ytaniza@ncsu.edu}

\affiliation{
Department of Physics, North Carolina State University, Raleigh, NC 27607 USA
}
\affiliation{
RIKEN BNL Research Center, Brookhaven National Laboratory, Upton, NY 11973 USA
}

\date{\today}

\begin{abstract}
Quark-hadron continuity is a scenario that hadronic matter is 
continuously connected to color superconductor 
without phase transitions as the baryon chemical potential increases. 
This scenario is based on Landau's classification of phases 
since they have the same symmetry breaking pattern. 
We address the question whether 
this continuity is true as 
quantum phases of matter, which requires the treatment beyond
Ginzburg-Landau description.
To examine the topological nature of color superconductor, 
we derive a dual effective theory for $U(1)$ Nambu-Goldstone (NG) bosons and
vortices of the color-flavor locked phase, and discuss the fate of emergent higher-form symmetries. 
The theory has the form of a topological $BF$ theory coupled to NG bosons, 
and fractional statistics of test quarks and vortices arises
as a result of an emergent $\mathbb Z_3$ two-form symmetry. 
We find that this symmetry cannot be spontaneously broken,
indicating that quark-hadron continuity is still a consistent scenario. 
\end{abstract}

\maketitle

%\tableofcontents

\section{Introduction}

%--------------------------------------------------------------------------------------

One of the most fundamental questions in nuclear physics is to
identify possible phases of quantum chromodynamics (QCD) \cite{Stephanov:2004wx, Fukushima:2010bq, Ren:2004nn, Alford:2007xm, Casalbuoni:2018haw}.
Understanding the phase structures at finite baryon densities
is relevant to the physics inside neutron stars and
has been of interest to nuclear and astrophysicists \cite{Lattimer:2004pg}. 
Hadronic matter is expected to exhibit nucleon superfluidity at finite
densities.
At very high densities, 
color superconductivity \cite{Barrois:1977xd, Bailin:1983bm} appears
with symmetric-pairing pattern among light three flavors,
called color-flavor locked (CFL) phase
\cite{Alford:1998mk}.
As classical many-body physics, phases of matter is classified by the
pattern of spontaneous symmetry breaking.
Based on this view, 
it is proposed that 
nucleon superfluidity and the CFL phase are connected
with smooth crossover,
since they have the same symmetry breaking
pattern: 
This is the {\it quark-hadron continuity}~\cite{Schafer:1998ef}.

The question we would like to address here is
whether this continuity holds beyond 
Ginzburg-Landau (GL) paradigm. 
Classification of quantum phases,
i.e. zero-temperature phases of quantum many-body systems, requires
beyond-GL description, because local order parameters cannot capture {\it topological order}. 
Importance of topology 
has been recognized in understanding gapped quantum phases~\cite{Wen:1989iv,
Wen200710}. 
Microscopic picture of topological order is given by long-range entanglement~\cite{PhysRevLett.96.110405, PhysRevLett.96.110404, Chen:2010gda}, and its low-energy description has spontaneously-broken higher-form global symmetry~\cite{Gaiotto:2014kfa}.   
An important consequence is that states with different topological
order cannot be continuously connected and there should be a quantum
phase transition between them. 
In recent years, the role of topology for gappless quantum systems is
also gradually taken into account, and it potentially has an impact for
understanding cuprate superconductors~\cite{Chatterjee:2017pqv,
Sachdev:2017hzd}. 

The quark-hadron continuity is recently examined in the 
presence of superfluid vortices~\cite{Alford:2018mqj, Chatterjee:2018nxe,
Cherman:2018jir}.
In the CFL phase, the minimal superfluid circulation of vortices 
is a fractional number $1/3$~\cite{Balachandran:2005ev,
Nakano:2007dr, Eto:2013hoa}.
In addition to $U(1)$ circulation, they also carry color holonomies.  
It is pointed out that this is a physical 
observable using color Wilson loops, and results in
fractional statistics between test quarks and vortices
\cite{Cherman:2018jir}. 
This has a certain similarity with topologically ordered phase in
condensed matter physics, which poses a doubt
on quark-hadron continuity as quantum phases of matter~\cite{Cherman:2018jir}. 

In this paper, we carefully examine the role of topology in the CFL
phase\footnote{Earlier works along this direction include \cite{Nishida:2010wr,
Fujiwara:2011za}, in which the effects of color holonomies are not
considered.
In Ref.~\cite{Alford:2010qf}, Aharonov-Bohm (AB) scattering of quark off color magnetic
fluxes is studied in the 2SC phase, in which vortices are not
topologically stable. 
In Ref.~\cite{Chatterjee:2015lbf}, scattering of color-neutral
particles off CFL vortices is discussed.
}.
We first derive the low-energy effective field theory of the CFL phase starting from the gauged GL model. 
It describes 
Nambu-Goldstone (NG) bosons associated with the breaking of $U(1)$
symmetry and superfluid vortices in a unified  
way. 
In particular, the effective theory correctly encodes 
the relation between the superfluid vortex and Wilson loop,
which is responsible for the fractional statistics of colored test
particles and vortices. 
We clarify that the $\mathbb{Z}_3$ fractional phase is a consequence
of an emergent $\mathbb{Z}_3$ two-form symmetry in the effective theory,
generated by color Wilson loops. The charged object under this symmetry
is nothing but the CFL vortices. 
We show that this emergent two-form symmetry cannot be spontaneously
broken, and thus emergent two-form gauge field is confined.
This means that the CFL phase has a trivial topological structure,
and we conclude that quark-hadron continuity scenario is alive also as quantum
phases of matter.

%-------------------------------------------------------------------------------------
\section{Symmetry of color flavor locking}

% 
%\subsection{Derivation of effective Abelian gauge theory}

We consider $3$-flavor QCD with degenerate quark masses.\footnote{The
effect of mass difference will be discussed later.}
The system has the global
symmetry,
\begin{equation}
{SU(3)_\rmf\times U(1)\over \mathbb{Z}_3\times \mathbb{Z}_3}, 
\end{equation}
where $SU(3)_\rmf$ is the vector-like flavor symmetry, $U(1)$ is the quark-number symmetry, and two $\mathbb{Z}_3$ factors in the denominators are introduced to remove the redundancies among $SU(3)_\rmf$, $U(1)$, and $SU(3)$ color gauge invariance~\cite{Shimizu:2017asf, Gaiotto:2017tne, Tanizaki:2018wtg}.  
In the presence of large chemical potential, quarks form the Fermi
surface with large Fermi momentum.
Since QCD has asymptotic freedom, the presence of this typical large
energy scale suggests that the system is weakly coupled and
semiclassical computation becomes reliable~\cite{Alford:1998mk}.
Within the one-gluon exchange, quark-quark interaction is attractive in
the anti-symmetric channel, indicating the Cooper instability of Fermi
surface.
Motivated by this observation, it is quite useful to introduce the
diquark operator $\Phi$ using the quark field $q$ by
\be
\Phi_{c_1f_1}=\ve_{c_1c_2 c_3} \ve_{f_1 f_2 f_3} (q_{c_2 f_2}^t\im
\gamma_0\gamma_2\gamma_5 q_{c_3 f_3}).
\ee
Here, $c_i$ and $f_i$ represent the color and flavor labels,
respectively. The diquark field $\Phi$ is in the anti-fundamental
representation for $SU(3)_\rmc$ color and $SU(3)_\rmf$ flavor symmetry,
and it has charge $2$ under $U(1)$.

Using this diquark field, the simplest effective Lagrangian is given by
the gauged Ginzburg-Landau model
\begin{equation}
 S={1\over 2g_{\mathrm{YM}}^2}|G|^2
 +{1\over 2}|(\diff + \im a_{SU(3)})\Phi|^2
 +V_{\mathrm{eff}}(\Phi^\dagger \Phi, \det(\Phi)),
\label{eq:gauged_GL_action}
\end{equation}
where $a_{SU(3)}$ is the $SU(3)_\rmc$ color gauge field, $G$ is its field
strength, and the effective potential $V_{\mathrm{eff}}$ depends only on
the color-singlet order parameters, $\Phi^\dagger \Phi$ and $\det(\Phi)$, and
$V_{\mathrm{eff}}$ has the symmetry $[SU(3)_\rmf\times U(1)]/[\mathbb{Z}_3\times
\mathbb{Z}_3]$.
For simplicity of discussion, we neglect the effect of absence of
Lorentz symmetry due to the chemical potential, but the extension will
be straightforward.
Let us now assume that $V_\mathrm{eff}$ has the minima at 
\begin{equation}
\Phi^\dagger \Phi=\Delta_0^2\bm{1}. 
\label{eq:vacuum_manifold}
\end{equation}
Taking the determinant of both sides, we get
$|\det\Phi|=\Delta_0^3$. In the gauge-invariant language~\cite{Cherman:2017tey, Tanizaki:2017mtm}, classical vacua break the
global symmetry spontaneously as
\begin{equation}
{SU(3)_\rmf\times U(1)\over \mathbb{Z}_3\times \mathbb{Z}_3}\to {SU(3)_\rmf\times \mathbb{Z}_6\over \mathbb{Z}_3\times \mathbb{Z}_3}={SU(3)\over \mathbb{Z}_3}\times \mathbb{Z}_2. 
\label{eq:SSB_pattern}
\end{equation}
Picking up a classical vacuum with $\det(\Phi)=\Delta_0^3$, we can fix
the gauge of $SU(3)$ color group so that
\begin{equation}
\Phi=\Delta_0\bm{1}. 
\end{equation}
Since $\Phi$ is in the bi-(anti-)fundamental representation of
$SU(3)_{\rmc}\times SU(3)_\rmf$, the symmetry breaking pattern in this
fixed gauge looks as
\begin{equation}
{SU(3)_\rmc\times SU(3)_\rmf\times U(1)\over \mathbb{Z}_3\times \mathbb{Z}_3}
 \to
 {SU(3)_{\rmc+\rmf}\times \mathbb{Z}_6\over \mathbb{Z}_3\times \mathbb{Z}_3}, 
\end{equation}
where $SU(3)_{\rmc+\rmf}$ is the diagonal subgroup of
$SU(3)_{\rmc}\times SU(3)_\rmf$.
This is why it is called color-flavor locking~\cite{Alford:1998mk}.

\section{
Derivation of a unified theory of $U(1)$ NG bosons and CFL vortices 
}

In the CFL phase, there are massless NG bosons associated with the
spontaneous breaking of $U(1)$ baryon number symmetry, and we can
construct the phenomenological Lagrangian by nonlinear realization.
Because of the quark masses, CFL pions are massive and they can be
neglected at low energies. 
Starting from the gauged GL theory with $SU(3)_\rmc$ color gauge group,
we derive the effective low-energy theory that satisfy this requirement. 

In order to correctly describe the possible low-energy excitations
including higher dimensional object, it is important to take into
account the topology of ground state manifold.
Here, we take the gauge so that the diquark field $\Phi$ is a diagonal matrix, 
\begin{equation}
\Phi=\Delta_0
\begin{pmatrix}
\rme^{\im \phi_1}&0&0\\
0& \rme^{\im\phi_2}&0\\
0&0&\rme^{\im \phi_3}
\end{pmatrix}, 
\label{eq:CFL_VEV_1}
\end{equation}
where $\phi_i$ is $2\pi$ periodic scalar fields. 
This realizes (\ref{eq:vacuum_manifold}), and hence indicates the symmetry breaking pattern (\ref{eq:SSB_pattern}). 
This choice of gauge is an analogue of maximal Abelian gauge in Yang-Mills theory with adjoint scalars~\cite{tHooft:1981bkw}. 
In this gauge fixing, the local gauge redundancy becomes the Cartan subgroup of $SU(3)_\rmc$,
\be
{U(1)_{\tau_3}\times U(1)_{\tau_8}\over \mathbb{Z}_2} \subset SU(3)_\rmc. 
\ee
Here, $U(1)_{\tau_3}$ and $U(1)_{\tau_8}$ are $U(1)$ groups generated by
$\tau_3=\diag[1,-1,0]$ and by $\tau_8=\diag[1,1,-2]$,
respectively. 
%\footnote{\changed{
%Let us make a technical comment here. 
%It is possible to choose a different basis of Cartan subgroup instead 
%of the current one, $\{\tau_3 , \tau_8\}$. 
%%With a a different choice, we will get a different $K$ matrix. 
%%In addition, the normalization condition 
%%(\ref{eq:constraint_gauge_field_1}, \ref{eq:constraint_gauge_field_2}), 
%%or equivalently set of physically observable operators (\ref{eq:physical-wilson-loops}) 
%%are modified. 
%Note that such a basis change does not affect correlation functions of physical operators. 
%}
%}. 
Since the rotations by $\pi$ in $U(1)_{\tau_3}$ and
$U(1)_{\tau_8}$ gives the same transformation matrix,
$\diag[\rme^{\im\pi},\rme^{\im\pi},1]$, the group structure is divided
by $\mathbb{Z}_2$.
Let us denote the corresponding $U(1)$ gauge fields by $a_3$ and $a_8$, and then the low-energy effective action (\ref{eq:gauged_GL_action}) becomes 
\begin{equation}
S={1\over 2 g_0^2}\left(|\diff \phi_1+a_3+a_8|^2+|\diff \phi_2-a_3+a_8|^2+|\diff \phi_3-2a_8|^2\right). 
\label{eq:gauged_GL_1}
\end{equation}
Here, we omit the kinetic term of gauge fields since they become heavy by Higgs mechanism, and $g_0=\Delta_0^{-1}$. 

Each scalar $\phi_i$ is not gauge invariant, and the only
gauge-invariant combination is
\begin{equation}
\varphi=\phi_1+\phi_2+\phi_3,
\end{equation}
and this corresponds to the NG boson associated with the spontaneous
breaking of $U(1)$ symmetry.
Another important remark is that each Wilson loop of $a_3$ and of $a_8$
is not observable since the gauge group is $[U(1)_{\tau_3}\times
U(1)_{\tau_8}]/\mathbb{Z}_2$ instead of $U(1)_{\tau_3}\times
U(1)_{\tau_8}$~\cite{Aharony:2013hda}. Observable Wilson lines are generated by
\begin{equation}
W_3(C)^2,\; W_8(C)^2,\; W_3(C)W_8(C), 
\label{eq:physical-wilson-loops}
\end{equation}
where $W_3(C)=\exp\left(\im \int_C a_3\right)$ and $W_8(C)=\exp\left(\im
\int_C a_8\right)$.
As a related fact, the normalization of gauge fields $a_3$, $a_8$ has to
be modified from canonical choice of $U(1)$ gauge fields as
\be
\int \diff a_3\in \pi \mathbb{Z},\; \int \diff a_8\in \pi \mathbb{Z}, 
\label{eq:constraint_gauge_field_1}
\ee 
with the constraint 
\be
\int \diff a_3=\int \diff a_8 \bmod 2\pi. 
\label{eq:constraint_gauge_field_2}
\ee

We are interested in the role of vortex configurations in the
CFL phase, and they are realized as the defect of the
scalar field in the gauged GL description.
For description of topological defects, it is convenient to take an
Abelian duality\footnote{
An dual action for the CFL phase in which gluons are also dualized is
studied in Ref.~\cite{Hirono:2010gq}. Here we take a dual of $U(1)$ NG
bosons only. 
}. 
As a preparation, let us derive~\cite{Banks:2010zn} the Abelian dual of
the following model in $4$ dimensions,
\begin{equation}
S={1\over 2g_0^2}(\diff \phi+k a)\wedge \star (\diff \phi+k a), 
\end{equation}
where $\phi$ is the $2\pi$ periodic scalar field, $a$ is the $U(1)$
gauge field, and $k\in\mathbb{Z}$ is the $U(1)$ charge.
We can rewrite this theory by introducing the $\mathbb{R}$-valued
$3$-form field $h$ as
\begin{equation}
S={g_0^2\over 8\pi^2}h\wedge \star h-{\im \over 2\pi}h\wedge (\diff \phi+k a). 
\end{equation}
Solving equation of motion of $h$, we get $h={2\pi\im \over g_0^2}\star (\diff \phi+ka)$, and obtain the original action by
substitution. Instead of integrating out $h$, we solve the equation of
motion of $\phi$ first, and then we obtain that
\begin{equation}
h=\diff b , 
\end{equation}
with $U(1)$ two-form gauge field $b$. The action becomes 
\begin{equation}
 S={g_0^2\over 8\pi^2}|\diff b|^2+\im{k\over 2\pi}b\wedge \diff a.
\end{equation}
This is the dual action of the Abelian Higgs model with charge $k$. 

Applying this procedure to the effective action (\ref{eq:gauged_GL_1})
for the CFL phase, we obtain
\begin{equation}
 S_{\rm eff}
  = {g_0^2\over 8\pi^2}\sum_{i=1}^{3}|\diff b_i|^2+{\im\over
 2\pi}\sum_{i=1}^{3}\sum_{A=3,8}K_{i A}\, b_i\wedge \diff a_A ,
\label{eq:eff-ac}
\end{equation}
where the matrix $K$ is given by
\begin{equation}
K=\begin{pmatrix}
1&1\\
-1& 1\\
0&-2
\end{pmatrix}. 
\end{equation}
This is the low-energy effective gauge theory describing NG boson, 
vortices, and color Wilson lines\footnote{Let us point out that the effective theory in (\ref{eq:eff-ac}) is $4$-dimensional field theory and thus it is different from the $5$-dimensional theory proposed in \cite{Cherman:2018jir}. }.
It has a structure of a topological $BF$ theory coupled with massless
NG bosons. 
General properties of this theory will be discussed elsewhere~\cite{Hirono:2019oup}.

\section{Fractional statistics and an emergent $2$-form symmetry}

The effective theory derived here encodes the relation between the color
holonomies and superfluid circulations. 
In the dual description, we can define the vortex operator as the Wilson
surface operator:
\begin{equation}
V_i(M_2)=\exp\left(\im \int_{M_2} b_i\right), 
\end{equation}
where $M_2$ is a vortex worldsheet. 
Using (\ref{eq:eff-ac}), one can show that
the braiding statistics between the vortex $V_i$ and test quarks $W_A$ is given by\footnote{The denominator necessary to cancel
non-topological contributions due to the coupling of vortices with
massless NG bosons.}
\begin{equation}
{ \left<V_i (M_2) W_A (C) \right>
  \over
  \left<V_i (M_2)\right>
 }
 = \exp \left[ \, 2\pi \im K^+_{Ai} \,\, {\rm link}(C, M_2)\right], 
\end{equation}
where $\mathrm{link}(C,M_2) \in \mathbb Z$ is the linking number of $C$
and $M_2$, 
and $K^+_{Ai}$ is the Moore-Penrose inverse of $K$, 
\begin{equation}
K^+=\begin{pmatrix}
{1\over 2}&-{1\over 2}&0\\
{1\over 6}&{1\over 6}&-{1\over3}
\end{pmatrix}. 
\end{equation}
Now, let us recall that the physical Wilson loops consist only of
$W_3^2$, $W_8^2$, and $W_3W_8$. We find that $W_3^2=1$, $W_8^2=(W_3W_8)^{-1}$, and 
\begin{equation}
{\langle V_i(M_2) W_8(C)^2\rangle \over \langle V_i(M_2)\rangle}=\exp\left({{2\pi\im\over3}\, \mathrm{link}(C,M_2)}\right). 
\label{eq:fractional-statistics}
\end{equation}
This reproduces the observation made in a recent
paper~\cite{Cherman:2018jir}. 
%
%The fractional statistics can be understood as
%a result of the emergence of
%a $\mathbb{Z}_3$ two-form symmetry \cite{Gaiotto:2014kfa} in the CFL
%phase. 
Equation (\ref{eq:fractional-statistics}) indicates 
the emergence of a $\mathbb{Z}_3$ two-form symmetry \cite{Gaiotto:2014kfa} in the CFL phase, where 
the generators are Wilson loops, $W_8^2$, and charged
objects are CFL vortices $V_i$. 
The explicit transformation of this two-form symmetry is given by 
\begin{equation}
b_1\mapsto b_1+{1\over 3}\lambda,\; b_2\mapsto b_2+{1\over 3}\lambda,\; b_3\mapsto b_3-{2\over 3}\lambda,  
\label{eq:cfl_2form_symmetry}
\end{equation}
where $\lambda$ is a flat two-form $U(1)$ connection with $\int_{M_2}\lambda\in 2\pi\mathbb{Z}$ for $\p M_2=\emptyset$. Under this transformation, the action changes as 
\be
\Delta S={\im\over 2\pi}\int \lambda\wedge (2\diff a_8)\in 2\pi \im \mathbb{Z}, 
\ee
using the fact that $\int \lambda\in 2\pi\mathbb{Z}$ and $\int \diff a_8\in\pi \mathbb{Z}$. Since $\exp(-\Delta S)=1$, we have confirmed that this two-form transformation is the symmetry. 
Note that there is no one-form symmetry for $a$,
unlike the case of the $BF$ theory with level $k$. 
This is because ${\rm dim (coker \,} K )\neq 0$,
which is equivalent to the existence of massless NG modes.

\section{Implication for quark-hadron continuity}

If the CFL is a superfluid phase
with topological order, 
there should be an emergent higher-form symmetry,
and it has to be spontaneously broken. 
We have seen that there exists an emergent $\mathbb{Z}_3$ two-form
symmetry, whose charged objects are CFL vortices, $V_i$.
However, these vortices show the logarithmic confinement,
and $\left<V_i\right>$ vanishes as vortex world-sheets become larger. 
This implies that the $\mathbb{Z}_3$ two-form symmetry is unbroken. 
Consequently, there is no deconfined topological excitation and 
the emergent two-form symmetry does not change the
topological structure of ground states. 
Therefore, 
it does not rule out the possibility that 
the CFL phase is continuously connected to the nucleon
superfluidity.

This can be further supported by a general theorem of quantum field
theory, without relying on the mean-field approximation.
Since the $U(1)$ symmetry is spontaneously broken, interaction of
low-energy Lagrangian should be written by the derivative of NG boson,
${1\over 2\pi}\diff \varphi={g_0^2\over 4\pi^2\im} \star \diff(b_1+b_2+b_3)$.
If the vortex fluctuation is heavy enough, then the
topological defect of $\varphi$ is negligible in the path integral,
and ${1\over 2\pi}\diff \varphi$ 
is conserved $U(1)$ current, generating the $U(1)$ two-form symmetry.
There is a subgroup $\mathbb Z_3 \subset U(1)$, which could be a
different symmetry from Eq.~(\ref{eq:cfl_2form_symmetry}). 
Incidentally, those two symmetries act in the same way on physical
observable $\exp(\im\int b_i)$ as $2 \pi \im /3$ phase rotations. 
A generalized version~\cite{Gaiotto:2014kfa} of Coleman-Mermin-Wagner
theorem~\cite{Coleman:1973ci, mermin1966absence} states that $U(1)$
$p$-form symmetry cannot be broken in less than or equal to $p+2$ dimension, and
thus $U(1)$ two-form symmetry cannot be broken in our $4$-dimensional
spacetime.
Consequently, its subgroup $\mathbb Z_3 \subset U(1)$ is unbroken.
Since this symmetry has the same order parameter as 
the emergent $\mathbb{Z}_3$ two-form symmetry,
it cannot be broken either in CFL phase. 
This suggests the \textit{quark-hadron continuity beyond Ginzburg-Landau paradigm}.

\section{Breaking $SU(3)_\rmf$ flavor symmetry}

Let us consider the effect of explicit $SU(3)_\rmf$ breaking. 
To see this, we assume that $V_\mathrm{eff}$ has the minimum at  $\Phi^\dagger \Phi=\mathrm{diag}(\Delta_1^2,\Delta_2^2,\Delta_3^2)$. 
After gauge fixing, the diquark field is
\begin{equation}
\Phi=
\begin{pmatrix}
\Delta_1\rme^{\im \phi_1}&0&0\\
0& \Delta_2\rme^{\im\phi_2}&0\\
0&0&\Delta_3\rme^{\im \phi_3}
\end{pmatrix}, 
\end{equation}
instead of (\ref{eq:CFL_VEV_1}) (see, e.g., \cite{Eto:2009tr}).
The absence of $SU(3)_\rmf$ symmetry is translated as
$\Delta_i\not=\Delta_j$ for different $i,j$. Correspondingly, the dual
effective action is changed as
\begin{equation}
 S_{\rm eff}
  ={1\over 8\pi^2}\sum_{i=1}^{3}g_i^2|\diff b_i|^2+{\im\over 2\pi}\sum_{i=1}^{3}\sum_{A=3,8}K_{i A}\, b_i\wedge \diff a_A, 
\end{equation}
with $g_i=1/\Delta_i$. 

To find the statistics, let us consider the equation of motion under the
presence of $V_3(M_2)$ vortex, which again has $1/3$ circulation. Equations of motion of $a_3$, $a_8$ are 
\begin{equation}
\diff b_1=\diff b_2=\diff b_3. 
\end{equation}
Equations of motion of $b_1,b_2,b_3$ say 
\begin{eqnarray}
 &&{g_1^2\over 4\pi^2}\diff\star \diff b_1
  ={\im\over 2\pi}\diff(a_3+a_8), \nonumber\\
 &&{g_2^2\over 4\pi^2}\diff\star \diff b_2
  ={\im\over 2\pi}\diff(-a_3+a_8), \nonumber\\
&&{g_3^2\over 4\pi^2}\diff\star \diff b_3
  ={\im\over 2\pi}\diff(-2a_8)-\im \delta^{\perp}(M_2). 
\end{eqnarray}
where $\delta^{\perp}(M_2)$ is the two-form valued delta function whose
support is $M_2$. 
As a result, for example, we find 
\begin{equation}
 \frac{
 \left<
  V_3(M_2)
 W_8(C)^2 
\right>
}{\left<  V_3(M_2)\right> }
=\exp\left({2\pi\im g_3^2\over g_1^2+g_2^2+g_3^2}\, {\rm link}(C,M_2)
     \right), 
\end{equation}
which is not quantized to $\mathbb{Z}_3$ phase unless we require $g_1=g_2=g_3$ coming out of $SU(3)$ flavor symmetry. 
In the absence of $SU(3)_\rmf$ symmetry, two-form symmetry generated by Wilson loops becomes an infinite group, in general. 
Since this may be regarded approximately as $U(1)$ two-form symmetry, the vortices should be confined by generalized Coleman-Mermin-Wagner theorem.

\section{Summary and conclusions}

We have derived the effective gauge theory
of CFL phase describing the NG bosons and vortices. 
The fractional statistics between vortices and colored test particles  
is shown to be a result of an emergent $\mathbb{Z}_3$ two-form symmetry. 
Color Wilson loops are the
generator of symmetry, and the charged
objects are superfluid vortices.
This emergent two-form symmetry is unbroken since the vortex-vortex
interaction shows logarithmic confinement. 
This is also supported by the generalized Coleman-Mermin-Wagner
theorem since we can find $\mathbb{Z}_3$ two-form symmetry is a subgroup
of emergent $U(1)$ two-form symmetry generated by ${1\over 2\pi}\diff
\varphi$.
Therefore, the symmetry breaking pattern of the CFL phase is the same
as that of nucleon superfluidity not only for ordinary symmetries but
also for higher-form symmetries. 
The effect of explicit $SU(3)_\rmf$ breaking is also studied, and we
checked that no higher-form symmetry is spontaneously broken.
Our analysis indicates that
the quark-hadron continuity scenario is consistent also as quantum
phases of matter. 

Our analysis suggests that there is some continuous local deformation of QCD Hamiltonian at finite densities that connects hadronic superfluid and CFL phase without quantum phase transition. 
It is important to point out, however, that we do not know if the chemical potential direction corresponds to this continuous deformation, so there may exist phase transition when we change the baryon chemical potential. 
Answer for this question requires the knowledge on dynamics of finite-density QCD, and one must go beyond the kinematical approach based on symmetry, anomaly matching, etc. 

Lastly, let us make several comments. 
The current work is based on a Lagrangian in the mean field
approximation, however whole analysis is translated into the language of
generalized global symmetry.
This indicates that the result of our analysis does not change under the
effect of perturbative fluctuations.
Vortices can appear as excited states (by rotation, for example). 
There are Majorana-fermionic excitations inside them 
\cite{Yasui:2010yw, Fujiwara:2011za, Yasui:2010yh,Hirono:2012ad}.
Roles and consequences of possible physics from those
states inside neutron stars are to be understood.

\begin{acknowledgements}
Y.~T. appreciates useful discussions with Aleksey Cherman and Srimoyee
 Sen. 
The work of Y.~T. was partly supported by Special Postdoctoral Researcher Program of RIKEN, and also by JSPS Overseas Fellowship. 
The work of Y.~H. was supported in part by the
Korean Ministry of Education, Science and Technology,
Gyeongsangbuk-do and Pohang City for Independent
Junior Research Groups at the Asia Pacific Center for
Theoretical Physics. 
\end{acknowledgements}

\bibliography{./refs,./QFT}

\end{document}